\documentclass[12pt]{article}
\usepackage{amssymb,amsmath,esint,titlesec}
\titleformat*{\section}{\normalsize\bf}
\titleformat*{\subsection}{\small\bf}

\begin{document}


\begin{titlepage}

\setlength{\baselineskip}{18pt}

                               \vspace*{0mm}

                             \begin{center}

{\Large\bf Irreversibility from staircases in symplectic embeddings}

                            \vspace*{25mm}

             \large\sf ANTHONY  \  \   J. \ \   CREACO \\
 
                           \vspace{2mm}           

        \normalsize\sf      Science Department, \\
                BMCC - The City University of New York, \\
             199 Chambers St., New York, NY 10007, USA.\\
                      acreaco@bmcc.cuny.edu\\
                         
                                   \vspace{15mm}

              \large\sf  NIKOLAOS \  KALOGEROPOULOS $^\dagger$\\

                            \vspace{2mm}
                            
 \normalsize\sf Center for Research and Applications \\
                                  of Nonlinear Systems \ (CRANS),    \\
   University of Patras, \  Patras 26500, \ Greece.                        \\
                    nikos.physikos@gmail.com \\
                         
                                    \end{center}

                            \vspace{25mm}

                     \centerline{\normalsize\bf Abstract}
                     
                           \vspace{3mm}
                     
\normalsize\rm\setlength{\baselineskip}{18pt} 

We present an argument whose goal is to trace the origin of the macroscopically irreversible behavior
of Hamitonian systems of many degrees of freedom.
We use recent flexibility and rigidity results of symplectic embeddings, quantified via the (stabilized)
Fibonacci and Pell staircases, to encode the underlying breadth of the possible initial conditions, 
which alongside the multitude of degrees of freeedom of the underlying system 
give rise to time-irreversibility.\\    

                           \vfill

                                                                         
                             \vfill

\noindent\rule{7cm}{0.2mm}\\  
   \noindent   $^\dagger$ {\footnotesize\rm Corresponding author}\\

\end{titlepage}
 

                                                                                \newpage                 

\rm\normalsize
\setlength{\baselineskip}{18pt}

\section{Introduction}

The origin of the macroscopic (time-) irreversibility observed in nature has been a topic of recurrent interest in Physics, since the late 19th century. 
The question it addresses is how does the macroscopically irreversible behavior arise, even though the underlying dynamics, in its Newtonian, Lagrangian, Hamiltonian
etc formulations, is time reversible. We are interested in classical, as opposed to quantum, behavior in this work, and we have in mind  particle systems, even though 
a large part of this discussion can presumably be carried over to the Statistical Mechanics of  fields. \\
  
The essence of the argument for irreversibility has been captured by L. Boltzmann's proposals \cite{Leb1, Leb2} who, in our opinion, has laid out the main ideas 
that lead toward a resolution of this issue. However deep Boltzmann's 
arguments are, they have always been considered heuristic, waiting for a more rigorous justification, which the hope was, ergodic theory might be able to provide.
This task is still unfinished though. The apperance of recent work on this issue such as \cite{Zak, CCCV} and the misunderstandings as pointed out in \cite{CCCV}
 in its proposed resolution point out toward its incomplete state of affairs, even today, and at the same time toward the conceptual and technical depth of this issue.\\ 

The relatively recent construction of a multitude of entropic functionals \cite{Ts-book}, and an independent but concurrent  re-examination of the foundations of Statistical 
Mechanics \cite{Touchette}, especially in the context of long-range interactions \cite{GuptaR},  is an additional  incentive for a parallel investigation of the arguments on the origin 
of time- irreversibility, especially in the broader context of the dynamical foundations of the ``non-classical", namely not the  Boltzmann/Gibbs/Shannon (BGS)  entropies.\\ 

In the conjectured dynamical framework of power-law entropies, we assumed in some of our previous works \cite{NK1, NK2, NK3, NK4}  that the non-additive behavior of systems described 
by the  $q$-entropy (also known as ``Tsallis entropy") may become manifest even for very few degrees of freedom. In the present work, we examine whether 
the time-irreversibility of Hamiltonian systems of many degrees of freedom can have any manifestation, even when one examines reduced systems of two effective degrees of  freedom. 
This is a level of reductionism that is exactly opposite to the complex systems that the $q$-entropy claims to describe, but it may be more technically tractable and may provide some form
of insight for the general case of  Hamiltonian systems of many degrees of freedom, whose description  is our ultimate goal.\\

In the present work, we do not use any particular entropic functional in our arguments.  We rely, instead, on the underlying dynamical description of  systems which are assumed to 
be modelled by autonomous Hamiltonians \cite{Arnold, AM, HZ}. We rely on results mostly obtained during the current decade, some of the references for which are   
\cite{Biran1, Biran2, Schlenk1, CHLS, Guth, McD1, McD2, BH1, McDSchl, BH2, HK1, PNg, Hind, FM, TPB, CGFSchl, CGH, CGHMcD, McD3,  Schlenk2}.
The present work can be seen as a physical application and interpretation of these results of symplectic geometry  to aspects of Hamitonian mechanics which may be pertinent to, 
and with a view toward, Statistical Physics.  Many of the above results in symplectic geometry  rely on and extend the foundational work of Gromov \cite{Gromov}
Hofer-Zehnder, Ekeland-Hofer \cite{HZ} etc, some of which were used in the closely related \cite{NK5} to argue for time irreversibility from essentially the  same  perspective.
 The current work extends \cite{NK5} which considered time irreversibility as a consequence of the symplectic  non-squeezing theorem, but 
 relies on the more recent mathematical developments stated  in the above references. 
Unlike the afore-mentioned  mathematical works which are rigorous, our arguments are hand-waving, attempting to provide a suggestive picture that may be pursued further 
in concrete models of physical significance, rather than firmly establishing generically applicable results.\\
   
 Our conclusion in the current work  is that the intricate pattern of flexibility and rigidity of symplectic embeddings
quantified through the stabilized symplectic staircases of ellipsoids into balls, which express the behavior of sets of initial conditions of the symplectic flows, 
alongside the large number of degrees of freedom of the full/unreduced systems,  can provide a plausible explanation for time-irreversibility, traces of  which can be detected 
even in systems having two effective degrees of freedom.\\ 

In Section 2, we provide some background from symplectic geometry in order to make the presentation reasonably self-contained to our intended audience, and to set up the notation.
Section 3 covers they key results  about the symplectic non-squeezing theorem and symplectic capacities. 
Section 4 contains recent results from the literature on symplectic embeddings and their obstructions. In Section 5, we point out how the above concepts and results can 
be interpreted as the source of time-irreversibility for Hamiltonian systems and manifest themselves for systems having two effective degrees of freedom. 
Section 6 states some conclusions  and posits some questions of interest to be tackled in the future.\\


\section{Symplectic basics}

There are several excellent sources for symplectic geometry and Hamiltonian dynamics these days. We have found parts of \cite{Arnold, HZ, ArnGiv, McDSal} to be very useful 
for foundational material. Moreover we have found the lucid and accessible work of M. de Gosson and collaborators \cite{DeG1, DeG2, DeG3, DeG4, DeG5, DeG6} on the symplectic view of 
classical, semi-classical  and quantum Physics to be a great motivation and help for this work. In addition, one may wish to consult \cite{MSB, SGM} for applications of the symplectic 
non-squeezing theorem in Astrophysics and Celestial Mechanics. These papers are also very enlightening in making quite transparent and 
concrete some otherwise abstract concepts of symplectic geometry and in smoothing out  terse aspects of some mathematical expositions.  One might add to this their considerable 
originality and the potentially substantial impact of their results for many more branches of Physics. \\ 


\subsection{Hamiltonian motivation}

We consider physical systems which are modelled by autonomous (time-independent) Hamiltonians \
 $H(x_1,\ldots, x_n, y_1, \ldots, y_n): \  \mathcal{M} \rightarrow \mathbb{R}$ 
\ where \  $(x_i, y_i), \  i=1,\ldots, n$ \ are  (Hamiltonian-) conjugate coordinates parametrizing the  $2n$-dimensional  phase space \ $\mathcal{M}$ \ of the system.
Without loss of generality, we will 
assume that \ $\mathcal{M} = \mathbb{R}^{2n}$, \ at least at this initial stage. The evolution of the system, in time \ $t$, \  is described by Hamilton's equations
\begin{equation}   
         \frac{dx_i}{dt} \ = \ \frac{\partial H}{\partial y_i}, \hspace{15mm} \frac{dy_i}{dt} \ = \ - \frac{\partial H}{\partial x_i}, \hspace{15mm} i = 1,\ldots, n
\end{equation}
which can be rewritten, using the shorthand notation 
\begin{equation}
 z \ = \ (x_1, \ldots, x_n, y_1, \ldots, y_n)
\end{equation}
as 
\begin{equation}
        \frac{dz}{dt} \ = \ - J_0 \  \nabla H (z)
\end{equation}
where the \ $2n\times 2n$ matrix $J_0$ \ has the form
\begin{equation}
     J_0 \ = \ \left(    
                   \begin{array}{cc}
                          {0}_{n\times n} & -1_{n\times n}\\
                          1_{n\times n} &   0_{n\times n} 
                  \end{array}
                                                            \right)
\end{equation} 
where \ $0_{n\times n}$ \ and \ $1_{n\times n}$ \ stand for the zero and the unit (diagonal) \ $n\times n$ \ matrices. One can readily see that \ $J_0$ \ is antisymmetric, 
that \ $J_0^2 = -1_{2n\times 2n}$, \  and that \ $J_0$ \ rotates the coordinates of each 2-plane of canonically conjugate variables counterclockwise by \ $\pi/2$. \  
The vector field
\begin{equation}
     X_H \ = \ - J_0 \ \nabla H
\end{equation}
is called Hamiltonian vector field and is an element of the tangent bundle \ $T\mathbb{R}^{2n} = \mathbb{R}^{2n}\times \mathbb{R}^{2n}$. 
With this definition, the integral curves of \ $X_H$ \ determine the evolution of the system whose Hamiltonian is \ $H$, \ according to (5). 
Moreover, and based on its geometric interpretation, one can readily see the Hamiltonian vector field is tangent to the constant energy hypersurface $H:$ constant in phase space.
Another way of expressing the properties of  (5)  is by introducing the antisymmetric 2-form \ $\omega_0$ \ on the cotangent bundle 
\ $ T^\ast \mathbb{R}^{2n} = \mathbb{R}^{2n} \times \mathbb{R}^{2n} $ \   which is  given in the coordinate system induced by (2), as       
\begin{equation} 
     \omega_0  \ = \ \sum_{i=1}^n \ dx_i \wedge dy_i
\end{equation}
where \ $d$ \ stands for the exterior derivative. We  observe that \ $\omega_0$  \ is a closed form 
\begin{equation}
       d\omega_0 \ = \ 0
\end{equation}
and that it is non-degenerate, namely that for vector fields $X, Y \in T\mathbb{R}^{2n}$
\begin{equation}
        \omega_0 (X,Y) = 0 , \ \  \mathrm{for \ all} \ Y \  \implies \  X \ = \ 0 
\end{equation}
The non-degeneracy condition (8) establishes an isomoprohism between the tangent and cotangent bundles of \ $\mathbb{R}^{2n}$, \  or of  any symplectic manifold \  $\mathcal{M}$. \  
In going from \ $\mathcal{M} \ = \ \mathbb{R}^{2n}$ \ to any symplectic manifold \ $\mathcal{M}$, \ one generalizes the above constructions, in analogy with numerous other well-known 
geometric and topological constructions,  by assuming that it is only locally valid in a chart and then appropriately ``sewing" the charts together via an approrpiate class of transition functions.
Having considered all the above, the definition of symplectic manifolds becomes less opaque: a pair \ $(\mathcal{M}, \omega)$ \ comprised of a manifold \ $\mathcal{M}$ \ and  a
two-from \ $\omega$ \ is called symplectic,  if \ $\omega$ \ is closed and non-degenerate. \\


\subsection{Definition and interpretation}  

The definition of a symplectic manifold \ $(M, \omega)$ \ can be re-phrased in a more intuitive/synthetic way as follows. Consider  a closed curve \ $\gamma: [0,1] \rightarrow \mathcal{M}, 
\ \gamma(0) = \gamma(1)$ \  in $\mathcal{M}$. 
The curve is allowed to be self-intersecting (immersed in \  $\mathcal{M}$).  Consider a surface \ $S$ \ such that its boundary is \  $\gamma$. \ Then \
$S = \cup_i S_i, \ \ i=1, \ldots, k$ \ which is generally the union of \ $k$ \ topological disks \ $S_k$ \ which are assumed to be  properly oriented. The relative orientation of \ $S$ \ and \ $\gamma$ \ 
is such that \ $\gamma$ \ is positively oriented when traversed counter-clockwise with respect to \ $S$ \ (the familiar in Physics ``right-hand rule").
 \  Let \ $A(S)$ \ indicate the (signed) 2-dimensional area of \ $S$ \ and let \ $pr_i$ \ be the projection  on the 2-plane spanned by \ $(x_i, y_i), \  i=1, \ldots, n$ \ taking into account the relative orientations. Then  \ $\mathcal{M}$ \ is symplectic if for all such $\gamma$
\begin{equation}   
      A(S) \ = \ \sum_{i=1} ^n   A(pr_i (S))
\end{equation}

It is clear from this definition that symplectic manifolds have a peculiarly two-dimensional structure. Since \ $\omega$\ is antisymmetric, the length of any vector is zero, 
and angles between vectors cannot be defined. It is an anisotropic geometry since the area of a projection of any curve on a 2-plane spanned by \ $(x_i, x_j), \ i, j = 1, \ \ldots, n, \  i\neq j$
 \ trivially vanishes. Due to the properties of the symplectic form, there are at least two classes of submanifolds of interest: symplectic and Lagrangians submanifolds. A submanifold 
$\mathcal{N} \subset \mathcal{M}$ \ is called symplectic, if \ $\omega$ \ restricted to \ $\mathcal{N}$ \ is symplectic \ $\omega|_{\mathcal{N}} = 0$. \ A submanifold \ 
$\mathcal{L} \subset \mathcal{M}$ \ is called Lagrangian, if \ $\omega|_{\mathcal{L}} = 0$ \ and it has maximal possible dimension which turns out to be half of that of \ $\mathcal{M}$, \ 
namely \ $\mathrm{dim}\mathcal{L} = n$.\\

Having worked out the above properties, one may wonder whether the transition and generalization  from Hamiltonian vector fields \ $X_H$,\ or their dual one-forms
\begin{equation}
    i_{X_H}\omega_0  \ =\ dH
\end{equation}
where \ $i$ \ in (10) indicates tensor contraction along a vector field, into symplectic manifolds may be too broad. The answer is essentially 
negative. One can easily see comparing (5) and (7) that this generalization amounts to extending the exact Hamiltonian one form into the closed symplectic  one. 
The Poincar\'{e}  lemma states that these two classes are equivalent, as long as the first cohomology group of \ $\mathcal{M}$ \ is trivial: \ \ $H^1 (\mathcal{M}, \mathbb{Z}) = 0$. \\

Two statements can be immediately made for Hamiltonian vector fields: the first is that the symplectic form \ $\omega$ \ is preserved under Hamiltonian flows, namely flows generated by 
Hamiltonian vector fields. To see this, apply Cartan's formula to the flow of \ $\omega$: 
\begin{equation} 
      \mathcal{L}_{X_H} \omega \ = \ d(i_{X_H}\omega) + i_{X_H}(d\omega) \ = \ 0
\end{equation}
In (11), \ $\mathcal{L}_{X_H}$ \ stands for the Lie derivative along the vector field \ $X_H$. \ Equation (11) is valid  because \ $d\omega = 0$ \  by definition, and since \ $X_H$ \ is a Hamiltonian vector field, then (11) implies that \ $d(i_{X_H}\omega) = ddH = 0$. \ As a result, the symplectic volume given by \ $\omega^n/n ! $ \ is preserved
\begin{equation}
                 \mathcal{L}_{X_H} \left( \frac{\omega^n}{n ! } \right) \ = \ 0 
\end{equation}
which is the familiar, from Classical Mechanics, Liouville's theorem. We notice that Liouville's theorem is a direct corollary of the preservation of the symplectic form \ $\omega$ \ under the Hamiltonian flow
generated by \ $X_H$. \  So a symplectic flow is volume preserving. An important question whether the converse were true was perplexing people for a few decades: it  was settled in the negative, by Gromov's symplectic non-squeezing theorem \cite{Gromov}, to be mentioned in the sequel. One can expand upon this non-equivalence and wonder  whether familiar  statements from Classical Mechanics can be implications of symplectic, rather than of  volume-preserving, geometry.
One such occasion, pertinent to Statistical Mechanics, would be to investigate the validity of the Poincar\'{e} recurrence theorem based on symplectic rather than on  volume preserving grounds 
\cite{MSchwarz, Schlenk2}. The issue remains largely open and subject to current investigation.\\
  
 Another  statement that we can make pertains 
to the geometric meaning of the condition (9). Consider a simple (oriented) closed curve \ $\gamma \in \mathcal{M}$ \ and let\ $D$ \ be its interior which is a topological disk, so that \ 
$\partial D = \gamma$. \ Then, using Stokes' theorem:      
\begin{equation}
       \int_D \omega  \ = \ \int_{\partial D} dH \ = \ \int_{\gamma} dH 
\end{equation}
Hence, if one deforms homotopically \ $D$, \  keeping its boundary curve fixed, then the integral will not change, which is a familiar condition in Physics \cite{Nak} signifying ``topological invariance".
Therefore,  symplectic geometry has a strong topological flavor, something that will become clearer in the sequel.\\ 


\subsection{Symplectic diffeomorphisms and Lagrangian submanifolds} 

A symplectomorphism, or symplectic diffeomorphism, is a diffeomorphism \ $\psi: \mathcal{M} \rightarrow \mathcal{M}$ \ 
which preserves the symplectic structure, namely for 
any two vector fields \ $X,Y \in T\mathcal{M}$
\begin{equation}
       (\psi^\ast\omega ) (X,Y) \ = \ \omega (\psi_* X, \psi_* Y)
\end{equation}
Since symplectomorphisms preserve the symplectic structure, in the more restricted contect of Hamiltonian mechanics they preserve the Hamiltonian, so they are the canonical transformations
familiar from Classical Mechanics. A symplectomorphism is obviously a volume-preserving diffeomorphism,  as can be seen in  (12). The question about the converse
 will be addressed  in the next Section.\\

Lagrangian submanifolds are intimately related to symplectomorphisms. This happens because of the following: Let  \ $\psi: (\mathcal{M}, \omega) \rightarrow (\mathcal{M}, \omega)$  \ 
be a  differomorpghism.  Consider the symplectic manifold \  $(\mathcal{M}\times\mathcal{M}, \omega\oplus (-\omega))$ \ and in it the graph of $\psi$: \ $\Gamma_{\psi} \subset  \mathcal{M}\times\mathcal{M}$ \ 
\begin{equation}
            \Gamma_\psi \ = \ \{ (w, \psi(w)), \ w\in \mathcal{M} \}
\end{equation}
Then, a theorem states that \ $\psi$  \ is a symplectomorphism if and only if the graph \ $\Gamma_{\psi}$ \ is a Lagrangian submanifold. As a result, the study of symplectomorphisms of 
a symplectic manifold is equivalent to the study of the Lagrangian submanifolds in its Cartesian product. \\  

Topology is at the core of symplectic geometry. The term ``symplectic topology" is actually a more accurate characterization of the content of the discipline.
This can be seen in a variety of ways, the most common of which is Darboux's theorem.
The theorem states that up to symplectomorphisms all symplectic manifolds have locally the structure of (6).
Hence it is impossible to distinguish locally between any two such manifolds having the same dimension: 
the differences between two equi-dimensional symplectic manifolds, if any, only arise at the global (topological) level.
This is in very sharp contrast to the Riemannian case, for instance, where the  Riemann tensor is a local invariant which is used to 
distingush the metric properties between two manifolds. This distinction is intimately related to the fact that the group of symplectomorphisms of a symplectic manifold is infinite dimensional,
in sharp contrast to the group of isometries of a Riemannian manifold which is not only finite dimensional, but its dimension is  (in most cases) fairly small.\\
     

\section{The non-squeezing theorem and symplectic capacities}

Symplectic diffeomorphisms are a core feature of symplectic geometry. Hence understanding their properties, or more concretely properties of the group of symplectomorphisms of a symplectic 
manifold, (or, in more physical terms, understanding the importance of canonical transformations of the phase space of a system)  is of central importance \cite{Polt-book}.\\


\subsection{The Gromov alternative}

 A particular question which arose during the 1960s, was the determination of the difference, if any, between the group of  symplectomorphisms \ $\mathsf{Symp}(\mathcal{M}, \omega)$ \  and the 
group of volume-preserving diffeomorphisms \ $\mathsf{VolDiff}(\mathcal{M}, \omega)$ \   of a symplectic manifold \ ($\mathcal{M}, \omega$). \  Let \ $\mathsf{Diff}(\mathcal{M})$ \  
stand for the group of diffeomorphisms of the differentiable manifold \ $\mathcal{M}$. \ 
One of the first notable results in this direction was the ``Gromov alternative" which was formulated in the early 1970s \cite{Gromov-book}:
consider the symplectic manifold \ ($\mathbb{R}^{2n}, \omega_0$). \ Then \
$\mathsf{Symp}(\mathbb{R}^{2n}, \omega_0)$ \ is either closed in \ $\mathsf{Diff}(\mathbb{R}^{2n})$ \ in the uniform ($C^0$-) topology (``rigidity"/``hardness"), or its \ $C^0$- closure is the 
whole group \ $\mathsf{VolDiff} (\mathbb{R}^{2n})$\ (``flexibility"/ ``softness"). \  This is one of the earliest indications of the flexibility and rigidity aspects (``soft" and ``hard" in the terminology of M. Gromov) of \ $\mathsf{Symp}(\mathcal{M}, \omega)$. \ Its resolution in favor of the rigidity result, namely that \ $\mathsf{Symp}(\mathbb{R}^{2n}, \omega_0)$ \ is \  $C^0$-closed in 
\ $\mathsf{Diff}(\mathbb{R}^{2n})$ \  was established by a combination of the work of M. Gromov \cite{Gromov, Gromov-book} and  Ya. Eliashberg \cite{Eliashb1, Eliashb2} from the late 1970s to the mid-1980s. The result means that the limit, in the uniform topology, of elements of 
$\mathsf{Symp}(\mathbb{R}^{2n}, \omega_0)$  is an element of $\mathsf{Symp}(\mathbb{R}^{2n}, \omega_0)$, \  and  that any element of \ $\mathsf{Symp}(\mathbb{R}^{2n}, \omega_0)$ \
 can be seen as the limit of a sequence of elements of \ $\mathsf{Symp}(\mathbb{R}^{2n}, \omega_0)$.   \\


\subsection{The symplectic non-squeezing theorem} 

A concrete  geometric manifestation of this rigidity was established in \cite{Gromov} which proved the following fundamental result, 
known as the ``symplectic non-squeezing theorem": Let \ $B_{2n}(r)$ \ be the closed ball in \ $\mathbb{R}^{2n}$ \  centered at the origin and of radius \ $r$
\begin{equation}
          B_{2n}(r) \ = \ \left\{  (x_1, \ldots, x_n,y_1,\ldots,y_n) \in \mathbb{R}^{2n}: \  \  x_1^2+\ldots +x_n^2+y_1^2+\ldots +y_n^2 \ \leq \ r^2   \right\}
\end{equation}
and let \ $Z_{2n}(R)$ \ stand for the (symplectic) cylinder in \ $\mathbb{R}^{2n}$ \  whose base is a symplectic 2-disk centered at the origin of radius \ $R$ , \ which for concreteness let's assume it as 
lying in the \ ($x_1, y_1$) \  2-plane:
\begin{equation}
       Z_{2n}(R) \ = \ \left\{  (x_1,\ldots, x_n, y_1,\ldots, y_n) \in \mathbb{R}^{2n}: \ \  x_1^2 + y_1^2 \ \leq \  R^2 \right\} 
\end{equation} 
The non-squeezing theorem states that \ $B_{2n}(r)$ \ can be symplectically embedded in \ $Z_{2n}(R)$ \ if and only if \ $r \leq R$. \ The result is only true as long as one uses embeddings in symplectic cylinders. If one considers a cylinder with base any isotropic 2-plane, such as \ $(x_i, x_j), \ i,j = 1,\ldots, n$ \ or similarly for \ $(y_i, y_j), \  i,j =1, \ldots, n$ \  the non-squeezing theorem does not apply. \\

Despite its apparent simplicity, the symplectic non-squeezing has far-reaching consequences. First of all, and regrettably, there is no proof of it easily understood by a non-specialized audience, even today. Second, it established symplectic geometry as a distinct discipline from volume-preserving geometry. Third, the method of pseudo holomorphic curves  that was employed in \cite{Gromov} is highly geometric and has been the cornerstone of a whole line of investigations until today. Fourth, such methods have provided a fertile ground not only for the development of symplectic geometry, but also for its connections to Physics (Yang-Mills and String-Brane/M- Theory)  such as Floer homologies, Mirror symmetry, Fukaya Categories, Gromov-Witten invariants etc.\\   


\subsection{Digression: geometrization and non-perturbative aspects of field theories} 

A digression may be in order at this point: pseudo holomorphic curves are in symplectic geometry the analogues of geodesics in metric geometry. They have one complex dimension (hence the word ``curves"), therefore two real dimensions. So, they are the  minimal 2-dimensional surfaces over which the symplectic form \ $\omega$ \ can be integrated, hence their connection to 2-dimensional 
objects in Physics such as the string world-sheet. Gromov exploited the close relation between the almost complex structure \ $J_0$ \  (4)  and the symplectic structure \ $\omega_0$ \ of a symplectic manifold to translate the symplectic embedding problem into a geometric embedding of  2-dimensional surfaces into an almost complex manifold target space. The latter  may be interpreted as a spacetime with matter and additional space-time supersymmetry  generators which provide the assumed almost complex structure(s).
In a sense, what Gromov did is the opposite of some current approaches in Physics, such as Loop Gravity \cite{LG}, which attempt to translate a partly geometric problem, which is the quantization of the metric of General Relativity into a Yang-Mills problem.\\

The approach of Loop Gravity uses the metric structure of space-time, but only when needed and temporarily as a background, for the sections of the frame bundle
and the associated connections of appropriate fiber bundles, which are elevated to the primary objects of study. The advantage of this approach is that one can rely on 
prior extensive experience in  quantizing Yang-Mills theories. By contrast, the quantization of the metric following a conventional canonical or path-integral approach  is practically 
perturbative (in 4 space-time dimensions) and  has lead to 
non-renormalizable Lagrangian densities \cite{GS}, both of which are undesirable features in a quantum theory of gravity. \\

The pseudo holomorphic curves approach points toward the opposite. It suggests that it may be beneficial to translate a Yang-Mills problem into a geometric one, and attempt to 
deal with aspects of the latter in a non-perturbative manner by using results of Topology and Geometric Analysis. This seems to be possible in symplectic (and contact) geometry due to 
 lack of local structure expressed through Darboux's theorem. Since we do not attempt to quantize the underlying system in this work, this does not negate or contradict in any 
way the approach that Loop Gravity or other attempts at quantization of Gravity may have followed. This translation of a Yang-Mills into a metric problem may not be a novel idea, but Gromov's successful 
work in establishing the non-squeezing theorem  provides a hint that it may be worth re-visiting non-perturbative aspects of General Relativity and 
Yang-Mills theory by translating them into geometric ones. \\           


 \subsection{Symplectic capacities} 
                                        
Symplectic capacities are non-negative invariants of symplectic manifolds under symplectomorphisms, whose goal is to express  the ``symplectic size" of a symplectic manifold \ ($\mathcal{M}, \omega$). \
An example of  such capacity can be inferred from the non-squeezing theorem, and is occasionally called the ``Gromov width": it is the radius of the largest ball that can be symplectically embedded in \
($\mathcal{M}, \omega$). \ This is the symplectic analogue of the familiar concept of inscribed ball in Euclidean and Metric Geometry. The concept of symplectic capacities was introduced and formalized 
by I. Ekeland and H. Hofer, and subsequently extended to all symplectic manifolds by H. Hofer and E. Zehnder, see for instance \cite{HZ}, as follows. \ A  symplectic capacity 
\ $c: (\mathcal{M}, \omega) \rightarrow \mathbb{R}_+ \cup  \{ 0, +\infty \}$  \  is a non-negative number (or infinity) satisfying the following three properties:
\begin{itemize}
    \item {\sf Monotonicity:} \ let  \ $\psi: (\mathcal{M}_1, \omega_1) \rightarrow (\mathcal{M}_2, \omega_2)$ \  be a symplectic embedding.\ Then \ $c$ \ satisfies \\ 
                                                               \centerline{      $c(\mathcal{M}_1, \omega_1) \ \leq \ c(\mathcal{M}_2, \omega_2)$   }
    \item {\sf Conformality:} \ $c(\mathcal{M}, \lambda\omega) \ = \ \lambda^2  \ c(\mathcal{M}, \omega)$, \ for \  $\lambda \in \mathbb{R}$.
    \item {\sf Non-triviality:} \ for $B_{2n}(1)$ \ and \ $Z_{2n}(1)$ \ in \ $\mathbb{R}^{2n}$ \ as in (16), (17): \ $c(B_{2n}(1))  =  c(Z_{2n}(1))  =  \pi$. 
\end{itemize}

The existence of the symplectic capacities is guaranteed by the symplectic non-squeezing theorem. Their  explicit constructions and computations have been a non-trivial, and largely unfinished, task. 
Sets of totally different shapes and sizes may have the same symplectic capacity.  The only 
relatively well-known case has been the construction of symplectic capacities on manifolds \ ($\Sigma, \omega$) \ possibly with a boundary, having $n=1$ (namely  2  real dimensions, i.e. real surfaces).
According to the Siburg-Jiang theorem \cite{HZ} a symplectic capacity of such a surface  is the modulus of its total area (Lebesgue measure):
\begin{equation}
         c(\Sigma, \omega) \ = \ \big{|} \int_\Sigma \ \omega \ \big{|}
\end{equation}
In addition, for the case of an ellipsoid in \ $\mathbb{R}^{2n}$, \ it turns out that one can always bring it, after a symplectic transformation, to the form 
\ $E_{2n} (a_1, \ldots, a_n) \subset \mathbb{R}^{2n}$ \ where 
\begin{equation}
       E_{2n} (a_1,\ldots, a_n) \ = \ \left\{     (x_1, \ldots,x_n,y_1,\ldots,y_n) \in \mathbb{R}^{2n}: \  \ \pi  \sum_{i=1}^n   \frac{|x_i + \sqrt{-1} \ y_i|^2}{a_i} \ \leq \ 1 \right\} 
\end{equation}
and the vertical lines indicate the complex modulus. Moreover, assume that all cross-sectoinal areas 
 \ $a_i, \  i=1,\ldots,n$ \  have been arranged in ascending order \ $a_1 \leq \ldots \leq a_n$. \ Then
\begin{equation}
              c(E_{2n}(a_1, \ldots, a_n)) \ = \  a_1
\end{equation}
It should be noticed that due to the non-triviality property in their definition,  symplectic capacities  cannot be functions of the (symplectic) volume, except in the case of surfaces. 
Moreover, due to their general non-additivity, symplectic capacities are not even measures. The set of symplectic capacities on \  ($\mathcal{M}, \omega$) \ is convex, hence contractible, 
among which the Gromov width is the smallest of all capacities and the cylindrical width is the largest.   \\


\section{Symplectic embeddings and obstructions}      

Due largely to the influence and success of the non-squeezing theorem, there as been a flurry of activity in understanding the conditions and obstructions to 
symplectic embeddings. The significance of this approach for Physics is as follows: symplectic embeddings 
can be used to express the evolution of initial conditions of the system under study.
A particular Hamiltonian \ $H$, \ hence a Hamiltonian vector field \ $X_H$ \ are the starting point 
in trying to understand the evolution of a system. However, the existence of the canonical transformations, which are re-parametrizations preserving the form of 
Hamilton's equations, changes the form of the Hamiltonian, but still preserves the form of Hamilton's equations (1), which is what ultimately matters in the evolution of the system.
Due to the great difficulty of solving Hamilton's equations by using explicit parametrizations, it is desirable to be able to make general statements that do not depend on the
particular form of the Hamiltonian, but just on some of its coordinate-independent properties. \\  
     

\subsection{Initial conditions: physical motivation} 

There is  a variety of physically relevant initial conditions that may be of interest. It is impossible to make general, model-independent, statements because initial conditions may vary according 
to the theoretical needs  of the model and the experimental feasibility, as well as the goals, of each system under study.  
For statistical mechanical purposes and to keep technicalities minimal and make the calculations as  tractable as possible, 
we will focus on the following four classes of ``shapes" of inital  conditions. 
\begin{itemize}
   \item[$\circ$] {\sf Polydiscs:} This is the Cartesian product of 2-discs centered at the origin of initial conditions, with each 2-disk \ $D_i (a_i), \  i=1\ldots, n$ \ lying on a 2-plane of conjugate variables.
             These discs do  not necessarily have to have equal radii, or equivalently, areas of projections \ $a_i, \ i=1,\ldots, n$ \ onto the symplectic 2-planes \  ($x_i,y_i$) \  in phase space. 
            Here, we follow the same notation as in (19) for ellipsoids. 
            A polydisc is the Cartesian product: \ $P_{2n} (a_1, \ldots, a_n) = D_1 (a_1) \times \ldots \times D_n (a_n)$. \
            This appears to be the simplest and most ``equitable" set of initial conditions as each pair of canonically conjugate variables is treated independently, but in a 
             similar manner qualitatively, to all other pairs. This set of initial conditions resembles a set of \ $n$ \ independent harmonic perturbations in both position and canonical momenta, 
            subject to the constraint that their total deviation from the given system has constant ``energy" for each pair of canonical variables.            
 \item[$\circ$] {\sf Cubes:} These are the polydisks \ $C_{2n}(a) = P(a,\ldots,a)$. \  Cubes are a very special and highly symmetric case of polydiscs. 
 \item[$\circ$] {\sf Ellipsoids:} With the notation of ellipsoids of equation (19), this set of initial conditions allows all canonical variables to have a range of values subject to a rather mild contraint of 
            a given total deviation from the initial condition of interest.  In a sense, it is as if we have decoupled harmonic perturbations of various angular frequencies in a system where we only 
            impose the condition of their total energy being kept fixed.
 \item[$\circ$] {\sf Balls:} These are the ellipsoids \ $B_{2n}(a) = E_{2n}(a,\ldots,a)$. \ They refer to the case where all the harmonic deviations of the system in phase space have the same angular frequency. 
   \item[$\circ$] {\sf Stabilized initial conditions:} One can consider focusing not on the whole system, but on a small subset of its number of degrees of freedom let's say 
               \ $2k \ll 2(n+k), \  k\in\mathbb{N}$ \ of 
             them,  and try to follow their evolution, where the other variables just provide a background to which these \  $2k$ \  degrees of freedom are coupled. 
           This is typical in effective field theories, in renormalizaiton group calculations etc. The 
              constraint on the specific set of \ $2k$ \ degrees of freedom refers to this small subset and leaves all other \ $2n$ \ initial conditions free/unconstrained. 
              This is the case of a system which is placed in a thermostat, which has far greater spatial extent and heat capacity than the system under study. 
              This approach bears a strong resemblance to the set up leading to the canonical ensemble in equilibrium  Statistical Mechanics. In such a case, 
                  the initial conditions of the four previous cases are modified to an embedding in a $2(n+k)$-dimensional symplectic space as 
                     \begin{equation}
                         P_{2k}(a_1,\ldots,a_k) \times \mathbb{R}^{2n}, \ \  C_{2k}(a) \times \mathbb{R}^{2n}, \ \  E_{2k}(a_1,\ldots, a_k) \times\mathbb{R}^{2n}, \ \           
                                                 B_{2k}(a) \times \mathbb{R}^{2n}
                     \end{equation} 
             respectively. Initial conditions of the form (21) are appropriately called ``stabilized" in the symplectic geometry literature, following analogous situations in Geometry and Topology.
\end{itemize}


\subsection{Symplectic embeddings: conditions and obstructions}

Following the success of the symplectic  non-squeezing theorem, the question arose \cite{CHLS} on whether one could perform similar a analysis for symplectic 
embeddings between different objects. From a physical viewpoint, such embeddings describe the Hamitonian evolution of sets of initial conditions. \\

An obvious obstruction to a symplectic embedding \  $(\mathcal{M}_1, \omega_1) \hookrightarrow (\mathcal{M}_2, \omega_2)$ \ is the volume constraint, if \ 
\begin{equation}
   \mathsf{Vol} (\mathcal{M}_1, \omega_1) \ >   \  \mathsf{Vol} (\mathcal{M}_2, \omega_2) 
\end{equation}
due to Liouville's theorem, where \ $\mathsf{Vol}$ \  in (22) stands for ``symplectic volume". 
We are interested in embeddings of equi-dimensional manifolds. Immersions are not that interesting mathematically, since all of \ $\mathbb{R}^{2n}$ \  
can be symplectically immersed into an arbitrarily small ball \ $B_{2n}(r)$. \ From a Physics viewpoint, such immersions may be relevant if one starts from a Hamiltonian
for the whole ``universe" and tries to derive an effective Hamiltonian for the system under study. Since the symplectic immersion approach does not provide 
any new results, we will ignore this possibility in the sequel.\\

 Embeddings into manifolds of higher dimension do not provide any new results either,  due to the existence of an $h$-principle \cite{Gromov-book}
which shows that all such embeddings are ``flexible". So unless someone places additional conditions  upon such embeddings, 
such as being Lagrangian submanifolds for instance, see \cite{McDSal}, we will ignore this case too. On physical grounds the general case may not be all that 
interesting as it describes an embedding of the system under study into a larger one, but because Lagrangian submanifolds, such as the  space of configurations 
of physical systems, are of great physical importance, imposing the rather strong Lagrangian submanifold embedding constraint may provide some non-trivial results.    \\ 

It follows, from the symplectic non-squeezing theorem, that a ball \ $B_{2n}(r)$ \ can be symplectically embedded into another ball \ $B_{2n}(R)$ \ 
if and only if \ $r \leq R$. \    Beyond this example and the ones in the next subsections,  
very little is currently known about symplectic embeddings of even regular shapes into other such shapes.
What is known is mostly limited to 4-dimensional embeddings ($n=2$) since the low dimension combined with techniques developed based on 4-dimensional gauge theories are 
strong enough to provide some answers. For this reason, in the sequel we will confine our attention to embeddings of the 4-dimensional shapes of the previous subsection.   
A 4-dimensional ellipsoid \ $E(a_1, a_2)$ \ can always be brought into the form \ $E(1,a), \ a\geq 1$ \ under a symplectic transformation. As a result, and after an application of the 
non-squeezing theorem, and since  \  $B_4(1) \subset E_4 (1,a) \subset Z_4(R)$, \ we see that \ $E(1,a)$ \ can be symplectically embedded into the cylinder \ $Z_4(R)$, \  if \
$R \geq 1$. \ This is a manifestation of the substantial rigidity of the symplectic maps.\\   


\subsubsection{The Pell staircase}

Another embedding problem whose answer is  known \cite{FM} is to determine the conditions, and obstructions, for the embedding \ $E_4(1,a) \hookrightarrow C_4(A)$. \ 
To quantify this, \cite{FM} determined the function
\begin{equation} 
         c^{EC}_0 (x) \ = \  \inf \{ \mu : \  E_4 (1,x) \hookrightarrow C_4(\mu ), \  \    x\geq 1 \} 
\end{equation}
where the embedding is assumed to be symplectic.
One can see that due to the symplectic non-squeezing theorem, there is the volume constraint 
\begin{equation}
       c^{EC}_0 (x) \ \geq \   \sqrt{\frac{x}{2}}
\end{equation}
 To even formulate the answer of \cite{FM} we need some preliminary notation. The silver ratio, as is well-known since antiquity, is 
\begin{equation}
    \sigma \  =  \  1+ \sqrt{2}
\end{equation}
the Pell numbers \ $P_n\in \mathbb{N}$ \ and the half companion Pell numbers \  $H_n\in\mathbb{N}$ \ are  defined by the recursion relations:
\begin{equation}
           P_0 = 0,  \hspace{15mm} P_1  =  1, \hspace{15mm}  P_n = 2P_{n-1} + P_{n-2}
\end{equation}  
and
\begin{equation}
          H_0 = 1, \hspace{15mm} H_1   =  1, \hspace{15mm}  H_n = 2H_{n-1} + H_{n-2}
\end{equation}
 or, more directly, by 
\begin{equation}
   P_n \ = \ \frac{(1+\sqrt{2})^n - (1-\sqrt{2})^n}{2\sqrt{2}}
\end{equation}
and
\begin{equation}
   H_n \ = \ \frac{(1+\sqrt{2})^n + (1-\sqrt{2})^n}{2}
\end{equation}
respectively. Now, consider the sequence 
\begin{equation}
        \left(         \delta_1, \ \delta_2, \  \delta_3, \  \delta_4, \   \delta_5, \ldots  \right) \ \  = \  \ \left( \frac{P_1}{H_0}, \  \frac{H_2}{2P_1}, \ \frac{P_3}{H_2}, \    \frac{H_4}{2P_3}, 
                                              \  \frac{P_5}{H_4}, \ldots  \right)
\end{equation}
which converges to \  \ $\sigma/\sqrt{2}$. \  The Pell staircase is defined as the graph formed by alternating the horizontal segments \ $\delta_n$ \ and the linear functions
meeting the previous horizontal sections and the graph of the volume function \ $\sqrt{x/2}$, \ in the domain \ $[1, \sigma^2]$. \  Then the theorem of \cite{FM} pertinent 
to our purposes states the following:
\begin{itemize}
   \item[(i)] On the interval \ $[1, \sigma^2]$, \ the function \ $c^{EC}_0(x)$ \ is given by the Pell staircase.
   \item[(ii)] On the interval \ $[\sigma^2, 7\frac{1}{32}]$, \ \  $c^{EC}_0(x)$ \ is the volume constraint \ $c^{EC}_0(x) = \sqrt{x/2}$ \ except on seven disjoint intervals 
                                where it is a piecewise linear function. 
   \item[(iii)] For \ $x\geq 7\frac{1}{32}$, \ \ $c^{EC}_0(x) = \sqrt{x/2}$ \ is the volume constraint.
\end{itemize}
In part (ii) of the theorem, \cite{FM} determined the exact functional form, which however is not needed for the purposes of  the present work, so we will forego its statement.. 
We observe in this theorem that the function \ $c^{EC}_0(x)$ \ proves that there is an alternating ``flexibility" and ``rigidity" of the symplectic embeddings  in the interval 
\ $[1, \sigma^2]$, \ which become mostly ``flexible" in the interval \ $[\sigma^2, 7\frac{1}{32}]$ \ with some occasional ``rigidity", and eventually all symplectic maps 
become totally flexible in their entirety for \ $x \geq 7\frac{1}{32}$. \\ 


\subsubsection{The Fibonacci staircase}

There is a well-known, now, criterion for the embedding of a 4-dimensional ellipsoid into another one, conjectured by H. Hofer and proved by D. McDuff \cite{McD2}. However, 
as \cite{Guth} proved, this criterion cannot be extended even to 6-dimensional ellipsoids, so we will forego stating it. Instead, we will consider in this subsection 
the potentially generalizable, to higher dimensions, case of the embedding of a 4-dimensional ellipsoid \  $E_4(1,x)$ \ into a ball $B_4(\mu)$ of 2-dimensional cross-sectional 
area \  $\mu$ \cite{McDSchl}.\\ 

Consider the function \ $c^{EB}_0: [1, \infty) \rightarrow \mathbb{R}$, \ defined as
\begin{equation}
    c^{EB}_0(x) \ = \ \inf \{\mu : \ E_4(1,x) \hookrightarrow B_4(\mu) \}
\end{equation}
where the embedding is assumed to be symplectic, in complete analogy with (23). This is another function which quantifies the flexibility and rigidity of symplectic embeddings. 
One observation is that this function is non-decreasing and continuous. Moreover it has as lower bound the volume function, as in the case of \ $c^{EC}_0(x)$, \  so 
\begin{equation}   
          c^{EB}_0(x) \ \geq \  \sqrt{x}\ 
\end{equation}
We recall that the Fibonacci numbers are defined through the recursion relation
\begin{equation}
   f_0 = 0, \hspace{10mm} f_1 = 1, \hspace{10mm} f_{n+1} \ = \  f_n + f_{n-1}, \ \ \  n\geq 1
\end{equation}
Let 
\begin{equation}
g_n = f_{2n-1}, \ \ \ n\geq 1
\end{equation}
 indicate the sequence of odd-indexed Fibonacci numbers.  Set 
\begin{equation}
    a_n \ = \ \left(\frac{g_{n+1}}{g_n}  \right)^2,    \hspace{15mm}   b_n  \ = \ \frac{g_{n+2}}{g_n}
\end{equation}
Then 
\begin{equation}
     \ldots \  < \ a_n \ < \  b_n \ < \ a_{n+1} \ < \  b_{n+1} \ <   \  \ldots
\end{equation}
and 
\begin{equation}
     \lim_{n\rightarrow\infty} a_n \ = \ \lim_{n\rightarrow\infty} b_n \ = \ \tau^4
\end{equation}
where 
\begin{equation}
       \tau \ = \ \frac{1 + \sqrt{5}}{2}
\end{equation}
is the golden ratio. Using this terminology, the main theorem of \cite{McDSchl} states that 
\begin{itemize}
     \item[(i)] The function \ $c^{EB}_0(x) \ = \ x/\sqrt{a_n}$ \ for \ $x\in [a_n, b_n]$, \ and \ $c^{EB}_0(x)$ \  is a constant, having value
                     \  $\sqrt{a_{n+1}}$ \ on the interval \ $[b_n, a_{n+1}]$, \ for all \ $n\geq 0$.
     \item[(ii)] The function \ $c^{EB}_0(x) = \frac{x+1}{3}$ \ in the interval \ $[\tau^4, 7]$.
     \item[(iii)] There are a finite number of closed disjoint intervals in \ $[7, 8\frac{1}{36}]$ \ where \ $c^{EB}_0(x) = \sqrt{x}$ \ for \ $x>7$ \
                     and as long as does not belong to any of these finite number of  intervals. 
     \item[(iv)] The function \ $c^{EB}_0(x) =  \sqrt{x}$, \ for \ $x  >  8\frac{1}{36}$
\end{itemize}
The alternating behavior of \ $c^{EB}_0(x)$ \ between ``flexibility" and ``rigidity"  of such symplectic embeddings,  
especially in the interval \ $x\in [1, \tau^4]$, \ reverting to total flexibility in \
$x > 8\frac{1}{36}$, \  is the main reason for the time-irreversible behavior of such Hamiltonian systems, as we will argue in the next Section.  \\ 


\subsubsection{The stabilized Fibonacci staircase}             

Ideally, for the purposes of Statistical Mechanics, one would like to understand the conditions for the symplectic embedding of one high dimensional 
ellipsoid into another one of the same dimension. As this turns out to be too difficult, one would like to  investigate the more tractable problem of the embedding 
of a high dimensional ellipsoid into an equi-dimensional ball. However, even this problem seems to be intractable today. 
As such we will have to settle with less. So, we can use the above results and wonder
what would happen in the ``stabilized" case. This is the case in which both domain and range are equi-dimensional shapes of initial conditions in phase space, 
but each of them has the special form indicated in  (21). \\

The answer to this question was provided, in part but sufficiently satisfactorily for our purposes, 
 in \cite{CGH, CGHMcD} which relied on a series of works that took place mainly during the present dacade. 
For the case of \ $1\leq x \leq \tau^4$ \ one finds that the function
\begin{equation}
       c_k(x) \ = \  \inf   \{\mu:  \ E_4 (1,x) \times \mathbb{R}^{2k} \hookrightarrow B_4(\mu) \times \mathbb{R}^{2k} \}   
\end{equation}
obeys 
\begin{equation}
       c_k (x) \ = \ c_0 (x)
\end{equation}
What may be worth noticing in this result is that the function \ $c_k(x)$ \ is independent of \ $k$ \ hence it ``stabilizes", in the topological language.
It states that the flexibility and rigidity properties of symplectic embeddings do not really detect the dimension of the ambient space (degrees of freedom of the ``thermostat" i n the Gibbsian 
canonical ensemble language). In the range \ $x > \tau^4$ \ not much is known except for the fact that 
\begin{equation}
       c_k(x) \ \leq \ \frac{3x}{x+1}
\end{equation}
Moreover, due to the volume bound \ $c_0(x) > \sqrt{x}$ \ we also have \ $c_k(x) < c_0(x)$. \ The question of whether 
\begin{equation}
       c_k(x) \ = \ \frac{3x}{x+1}, \ \ \  \ \ \   k\geq 1
\end{equation}
for \ $x > \tau^4$ \ is still under investigation \cite{McD3}.\ These results also showed that the graph of \ $c_0(x), \ \ 1 \leq x \leq \tau^4$, \ which is piecewise linear,  
``bounces" between that of \ $\sqrt{x}$ \ and that of \ $3x/(x+1)$. \\   


\section{Symplectic embeddings, coarse-graining and irreversibility}

The potential physical significance of the above results of symplectic geometry is the following. In Statistical Physics, we have a system of many degrees of freedom. Let us
focus our attention to just two of these degrees of freedom and the corresponding initial conditions. Such sets of initial conditions can be arbitrary, as mentioned above. However
for technical reasons as well due to the Central Limit Theorem, we focus on sets  having the shape of balls and ellipsoids. These 4-dimensional ellipsoids in phase space can have any
size. One may choose the initial conditions of these two effective degrees of freedom to be an ellipsoid or polydisc of the same size for each one of them, 
or one of them can be, in principle, larger/smaller than the other. But as we see in the sequel symplectic sections of such ellipsoids or polydics cannot have very different sizes.   
In Statistical Physics, the approach of Gibbs is to consider all such sets of initial conditions for a particular Hamiltonian evolution 
and then average over them  to reach the values of the quantities of physical  interest (the canonical ensemble).\\


\subsection{Time irreversibility from staircases}
In the case of ellipsoids and balls in 4-dimensions, the apparence of the Fibonacci (and Pell) staircases show that  some of these shapes of initial conditions are flexible 
and some are quite rigid under symplectic embeddings. The form of such embeddings is alternating between flexibility and rigidity. 
Statistical averaging over such initial conditions makes one lose track of which initial conditions were in the inverse image of the ball under a short-time Hamiltonian evolution.
So running backward the time evolution of the system is unable to track the inverse images of the different initial conditions which have eventually evolved into 
the shape of a ball.  In essence these staircases are a natural form of coarse graining for the Hamiltonian evolution of the system.
Due to averaging, the succession of flexibile and rigid sympletic maps makes it impossible for someone to distinguish between the images of the
 initial conditions with absolute accuracy. The result is the lack of the macroscopic time reversibility of the system. \\


\subsection{The role of the large number of degrees of freedom}

The large number of degrees of freedom of the initial Hamiltonian system plays an important role in how ``typical" are the ellipsoidal shapes of the initial conditions that we focus on. 
One could have chosen any set of initial conditions that they might find experimentally feasible or theoretically desirable to perform such an analysis, as was pointed out before.  
It might be easiest theoretically to choose initial conditions forming a convex set on a neighborhood of the given system. If one uses the full Hamiltonian of the 
system rather than its reduced description by only two degrees of freedom, then such initial conditions, chosen independently and identically for each degree of freedom,
would converge to a (high-dimensional) Gaussian, according to the Central Limit Theorem. 
But such a Gaussian is indistinguishable in the limit of many degrees of freedom from a ball, according to the Maxwell/Borel/L\'{e}vy asymptotic estimate on the sections of balls 
and Gaussians \cite{Ball}. The physical significance of this well-known geometric result for coarse graining of Hamiltonian  systems
was pointed out in \cite{NK6}, and for time irreversibility  in \cite{NK5}.\\

As a result, any projection into the 4-dimensional subspace of initial conditions of the two effective degrees of freedom of such a set will typically have the shape of a ball. 
To account for variations due to the small number \ $n=2$ \ of the effective degrees of freedom and potential perturbations, it may be advisable to
consider somewhat deformed balls, namely ellipsoids. However such ellipsoids must have sections that 
are not too different from each other in order to resemble reasonably well balls. After all, large deviations from the shape of balls are exponentially suppressed according to the 
Central Limit Theorem.  Hence our focus on \ $x\in [1, \tau^4]$ \ namely in the Fibonacci and the stabilized Fibonacci staircases,
as well as in \ $[1, \sigma^2]$ \ namely  in the Pell staircases, may be reasonable on physical grounds.\\


\subsection{Symplectic packings}

The above issues are not unrelated to the case of symplectic packings \cite{Schlenk2}, with which they share common developent and features. For concreteness, let us consider packings of balls
 into a cube. In this paragraph only, the cube is ``geometric" \ $C_{2n} = [0,1]^{2n}\subset \mathbb{R}^{2n}$. \ The problem is to try to fill as much as possible of the volume of 
\ $C_{2n}$ \ by \  $k\in\mathbb{N}$ \ balls of equal radius \ $R$ \ using symplectic maps. 
To quantify the problem, let \ $\mathsf{Vol}$ \ stand for Volume, let \ $\bigsqcup$ \ stand for disjoint union of sets, and  define the symplectic packing number
\begin{equation}  
       p_{2n}(k) \ = \ \sup_R \left\{k \ \mathsf{Vol} B_{2n}(R) :  \bigsqcup_k  B_{2n}(R) \hookrightarrow C_{2n}   \right\}
\end{equation}
where the embeddings are assumed to be symplectic. Obviously \ $p_2(k) = 1$ \ since in 2-dimensions ($n=1$), ``symplectic" is equivalent to ``volcume preserving". For \ $n\geq 3$ \ these 
numbers are not known. What is known however is that there are constants \ $k_0(2n)$ \ such that \ $p_{2n}(k) =1$, \ for all \ $k \geq k_0(2n)$. \ This is total symplectic flexibility, exhibited in 
the Pell staircase for \ $x\geq 7\frac{1}{32}$ \ and in the Fibonacci and stabilized Fibonacci staircases for \ $x\geq 8\frac{1}{36}$. \ However it is the  range of \ $x$ \ below these values which is
of physical interest, and we expect that in such cases \ $p_{2n}(k) < 1$ \  based on preliminary results \cite{Schlenk2}. 
This amounts to using balls whose radii are \ $R<1$, \ but which are not too small, in our attempt to fill the cube \ $C_{2n}$. \\ 

If a large proportion of the cube is filled, and because of its high dimension if we cannot distinguish distances that are too small, then the whole cube appears to be fully covered 
by the symplectic images of these balls. In such a case it is impossible to 
trace back the sizes of the balls which fill  the cube, namely determine the size and the number of balls in the inverse image 
that almost fill the cube.  So, it is impossible to determine whether such a filling is due to either many but smaller, or larger but fewer, balls. 
Therefore, the behavior of the packing function   provides a natural 
coarse graining for the interior of cubes of high dimension. Similar things can be stated abour packings inside balls.
The statistical averaging over initial conditions of different sizes in phase space of the previous paragraphs,
has been  replaced in the case of packings by a coarse approximation to the interior of the filled cube. 
Both approaches are can be expressed  in probabilistic terms, as Boltzmann had proposed \cite{Leb1, Leb2}.    \\   
 

\subsection{Fibonacci numbers, the golden ratio and the KAM theorem}

The appearence of the Fibonacci sequence and the golden ratio in Subsection 4.2  above may be unexpected despite their widespread appearence in a multitude of
occasions \cite{Dunlap}. 
We would like to mention its potential connections in the context of the Kolmogorov-Arnold-Moser (KAM)  theorem, see \cite{Poschel} for a more recent introduction.
 As is well-known the KAM theorem describes the stability of 
the $n$-dimensional family of Lagrangian tori in the phase space of integrable Hamiltonian systems, under small Hamiltonian perturbations. In the context of the KAM theorem
tori whose frequencies have the golden ratio are the last ones which are  destroyed under such perturbations. This happens because the golden ratio is the ``most" irrational number in the 
following sense:  Consider the golden ratio's continued fraction expansion, which is a standard way of obtaining the best rational approximation of any number. It is given by 
\begin{equation}
       G \ = \ 1+ \cfrac{1}{1+ \cfrac{1}{1 + \cfrac{1}{1+\cfrac{1}{1+ \cdots}}}}
 \end{equation}
As is well-known for any \ $x\in\mathbb{R}$ \ there is a constant \ $C(x)$ \ which measures how hard it is to approximate by rational numbers. It is defined by
\begin{equation} 
      \liminf_{q\rightarrow \infty} \left| x-\frac{p}{q} \right| \  = \ \frac{C(x)}{q^2}, \hspace{10mm}  p,q \in \mathbb{Z} 
\end{equation}
It turns out that \ $C(x)$ \ is as big as possible when \ $x$ \ is the golden ratio. The question is whether the appearence of the golden ratio in the KAM theorem and in the 
(stabilized) Fibonacci staircase is accidental or there is a deeper connection between them  in some non-trivial way. 
It  may be worth re-examining aspects of the KAM theorem from the viewpoint of the Fibonacci sequence
appearing in the symplectic embeddings of ellipsoids into balls, which one could imagine, or even conjecture, that they may provide a non-perturbative generalization to some 
of the statements present in the KAM theory. \\  

In closing this Section, we notice that we do not have to consider the possible form that the entropy which describes the macroscopic behavior of the system may have.
 We have just assumed a Hamiltonian evolution, and the validity of the Central Limit Theorem for sets of initial conditions  of the underlying Hamiltonian. 
Regardless of the possible effective correlations of the underlying variables
which may dictate various forms of entropy to describe the macroscopic behavior of the system, as is usually assumed \cite{Ts-book} for power-law entropies, 
our results remain valid, as they do not rely on any such assumptions and are completely dynamical, at the microscopic level.\\
 

\section{Conclusions and outlook}

In this work we attempted to provide a hand-waving  justification for the macroscopic irreversibility of Hamiltonian systems having many degrees of freedom. We relied on recent results in symplectic geometry
which extend in non-trivial ways the symplectic non-squeezing theorem, and which were obtained mostly during the current decade. We argued that time irreversibility may appear even in a system
of two effective degrees of freedom. \\

We have used recent, but known, results of symplectic geometry. From a different viewpoint, our work can be seen as providing a general statistical mechanical interpretation of a few of these 
symplectic geometric results. Our arguments have been hand-waving, as many aspects of Hamiltonian systems of many degrees of freedom are not understood, 
and not too many generic patterns have emerged in the description of the evolution of such dynamical systems so far  that may allow us to make more concrete and physically relevent statements. 
We have, out of necessity, confined ourselves to sets of inital conditions that are highly symmetric such as ellipsoids, 
balls and polydiscs. Due to the large number of degrees of freedom of the underlying system, we argued that these choices may be more generic than one might initially believe.\\ 

This work is part of our attempt to explore the dynamical foundations of the non-BGS entropic functionals. Some of the symplectic  results we use here, may prove to be effective
 in exploring the microscopic origin of phase transitions, along the lines of the ``topological hypothesis", whose development however has used a more metric rather than symplectic approach \cite{Pettini, Kastner}. One could use  Floer homology \cite{AD} for instance,  to explore from a symplectic viewpoint some of the conjectured relations between the behavior of the critical points of 
a function in phase space and the appearence of phase transitions in the macrocscopic behavior of a system, completely sidestepping the metric approach \cite{Pettini, Kastner}.  
Notice that nowhere in the present work have we assumed any specific form of the entropic functional that may describe the macroscopic behavior of the system.\\
 
 The present line of work may also be a motivation for the investigation of specific high dimensional phase spaces such as high dimensional symplectic or  K\"{a}hler manifolds \cite{Ballmann}, 
for instance.  The latter
have enough rigidity, and at the same time contain a symplectic structure, so they are not only  good, but also analytically tractable, testing grounds as configuration spaces  of systems of many degrees  freedom. As has been known since the times of J.C. Maxwell and L. Boltzmann, the large number of degrees of freedom provide considerable 
simplifications in the study of a system, and it is exactly this simplification that allows Statistical Mechanics to provide  robust predictions which make equilibrium Thermodynamics so accurate.
Do such simplifications also exist, and if so,  can they be  used to describe high dimensional symplectic or K\"{a}hler manifolds that may be the configuration spaces of physical systems? Can such questions provide a hint toward unraveling the dynamical basis, if any, of power-law entropies? These are questions that might be worth pursuing in the near future.  \\


               \vspace{0mm}

\noindent{\bf Acknowledgement:}  \ \ We are grateful to the referee for his/her constructive criticism of the manuscript which helped improve the exposition, and also for pointing out to us 
references \cite{MSB, SGM} of which we were completely unaware at the time of development and writing of this manuscript. 
 The second author is  grateful to Professor Anastasios Bountis  for his support, without which  this work would have never been possible.  \\  





\end{document}